# Weyl–Cosserat Elasticity and Gravitational Memory: An Effective Microstructured Model of Spacetime

**David Izabel**

**Abstract**

We construct a mathematically controlled correspondence between the electric and magnetic parts of the Weyl tensor in vacuum general relativity and the kinematics of a micropolar (Cosserat) elastic medium. In this framework, gravitational memory is reinterpreted as the topological charge of an effective dislocation field in spacetime. The ordinary displacement memory corresponds to an edge dislocation characterized by a non-trivial Burgers vector, while spin memory corresponds to a screw-type defect associated with rotational mismatch. We formulate the correspondence explicitly, derive it from the Bianchi identities and the geodesic deviation equation, and construct an effective Lagrangian extension of Einstein–Cartan theory describing propagating torsion modes. The framework is shown to be an effective coarse-grained description rather than a modification of classical GR, and we discuss its observational viability.

---

## 1. Introduction

General Relativity (GR) describes gravitational waves in vacuum via the Weyl tensor $C_{abcd}$, which satisfies $R_{abcd} = C_{abcd}$ in the absence of matter. The detection of gravitational waves by LIGO and Virgo, and the confirmation of the gravitational memory effect [4,5], motivate a reconsideration of spacetime as possibly possessing an effective mesoscopic structure [10].

We propose a novel interpretation:

Gravitational memory is the topological remnant of a transient defect in an effective spacetime medium.

This analogy is developed through a controlled correspondence between:

1. The 3+1 decomposition of the Weyl tensor.

2. The kinematics of micropolar (Cosserat) elasticity [1,2].
3. Dislocation theory and the Burgers vector [3].
4. The geometric framework of torsion [7].

A crucial clarification is necessary: this work does not claim that GR fundamentally possesses torsion. Instead, torsion emerges as an *effective, coarse-grained descriptor* of the permanent deformation left behind by a gravitational wave, i.e., the memory effect.

## 2. Theoretical Background

### 2.1 Riemann–Ricci–Weyl decomposition (4D)

On any $n \geq 4$pseudo-Riemannian manifold $(M, g)$,

$$R_{\mu\nu\alpha\beta} = C_{\mu\nu\alpha\beta} + \frac{1}{n-2}\left(g_{\mu[\alpha}R_{\beta]\nu} - g_{\nu[\alpha}R_{\beta]\mu}\right) - \frac{R}{(n-1)(n-2)}\, g_{\mu[\alpha}g_{\beta]\nu}$$

Here, $C_{\mu\nu\alpha\beta}$is the traceless, conformally invariant part of curvature. In vacuum in General Relativity the Ricci tensor is null ($R_{\mu\nu} = 0$), the full curvature reduces to Weyl, and gravitational shear waves are pure Weyl.

### 2.2 Weyl as a $6 \times 6$operator on bivectors

In 4D, the space of bivectors $\Lambda^2 T_p^* M$is 6-dimensional. Ordering antisymmetric index pairs as $\{01, 02, 03, 23, 31, 12\}$, one views $C_{\mu\nu\alpha\beta}$as a symmetric linear operator $C_{AB}$on this 6D space (with the bivector metric $G_{AB}$), providing a spectral perspective underlying Petrov types (I, II, D, III, N, O).

### 2.3 Weyl Tensor in 3+1 Decomposition

Let $u^a$ be a geodesic, timelike congruence of observers. The Weyl tensor splits into its "electric" and "magnetic" parts [14], both symmetric, trace-free, and spatial:

$$E_{ab} = C_{acbd}u^c u^d, H_{ab} = \frac{1}{2}\epsilon_{ac}^{mn} C_{mnbd}u^c u^d$$

In vacuum, the Bianchi identities $\nabla^d C_{abcd} = 0$ yield a set of Maxwell-like propagation equations [8]:

$$\dot{E}_{\langle ab\rangle} = \text{curl } H_{ab}, \dot{H}_{\langle ab\rangle} = -\text{curl } E_{ab}$$

This hyperbolic system ensures propagation at the speed of light and forms the basis for our correspondence.The electrical part $E_{ab}$: represents the shear deformation. The Magnetic part $H_{ab}$: represents transverse rotation / frame-dragging. These two objects are transverse, traceless, exactly like pure shear

### 2.4 Elasticity stiffness as a 6×6 operator (classical Cauchy elasticity)

In 3D elasticity, minor symmetries $C_{ijkl} = C_{jikl} = C_{ijlk}$and major symmetry $C_{ijkl} = C_{klij}$allow one to represent $C_{ijkl}$as a symmetric $6 \times 6$matrix in Voigt notation, acting on the 6D space of symmetric strains. The energy density $E = \frac{1}{2} C_{ijkl}\varepsilon_{ij}\varepsilon_{kl}$ensures symmetry and positive-definiteness for stable materials; only 21 independent components remain in the general anisotropic case.

Before introducing micropolar kinematics, it is useful to recall the classical (Cauchy) elasticity framework, where each material point carries only a displacement field $u_i(x,t)$.So, in 3D linear elasticity, the constitutive law relates the symmetric Cauchy stress $\sigma_{ij}$to the symmetric strain $\varepsilon_{ij}$through the rank-4 stiffness tensor. The fundamental kinematic quantity is the small-strain tensor:

$$\varepsilon_{ij} = \frac{1}{2}(\partial_i u_j + \partial_j u_i)$$

which measures symmetric, purely translational distortion.
The Cauchy stress $\sigma_{ij}$is related to strain through the linear constitutive law:

$$\sigma_{ij} = C_{ijkl}\, \varepsilon_{kl}$$

The stiffness tensor $C_{ijkl}$satisfies the minor and major symmetries:

$$C_{ijkl} = C_{jikl} = C_{ijlk} = C_{klij}$$

which allows the representation of $C_{ijkl}$as a symmetric $6 \times 6$operator acting on the 6-dimensional space of symmetric strains. The stored elastic energy is :

$$W = \frac{1}{2} C_{ijkl} \, \varepsilon_{ij} \, \varepsilon_{kl}$$

is positive definite for stable materials, leaving 21 independent components in tis:ully anisotropic case.This classical structure $(u_i, \varepsilon_{ij}, C_{ijkl})$provides a natural reference against which the Cosserat kinematics will be compared.

### 2.5 What the literature *does not* say – state of the art

Authoritative treatments of the Weyl tensor (monographs, lecture notes) establish all of its algebraic properties, including its action as a self-adjoint linear operator on the 6D space of bivectors. Conversely, elasticity monographs describe extensively the properties of the stiffness tensor $C_{ijkl}$(symmetries, invariants, spectral decomposition), but do not relate it to the Weyl tensor.

Mathematical works on traceless $(2,2)$tensors analyze such objects abstractly, yet do not formulate a physics-level identification between the Weyl operator and the deviatoric part of the elasticity stiffness operator.

Hence the following gap arises:

No reference explicitly states that “Weyl = stiffness (deviatoric) operator acting on 2-forms.”

Exploring this correspondence and its physical implications is one of the goals of this paper.

### 2.6 Micropolar Elasticity (Cosserat Medium)

In a Cosserat medium, each material point carries a displacement field $u_i$and an independent microrotation $\phi_i$. The classical symmetric strain $\varepsilon_{ij}$is replaced by the distortion tensor

$$\beta_{ij} = \partial_j u_i + \varepsilon_{ijk} \, \phi_k$$

which contains both translational and rotational contributions. A new curvature-like measure, the wryness tensor,

$$\kappa_{ij} = \partial_j \phi_i$$

encodes spatial gradients of micro-rotation.

Geometric incompatibility of the distortion field produces a dislocation density

$$\alpha_{ij} = \varepsilon_{jkl} \, \partial_k \beta_{il}$$

whose flux yields the Burgers vector

$$b_i = \oint_\Gamma \beta_{ij} \, dx^j = \int_S \alpha_{ij} \, dS^j$$

Compared to classical elasticity, the Cosserat medium extends the kinematic space

$$(u_i, \varepsilon_{ij}) \longrightarrow (u_i, \phi_i, \beta_{ij}, \kappa_{ij})$$

and therefore the constitutive structure also enlarges, enabling couple-stresses and torsional effects absent in Cauchy elasticity.

This structure generalizes classical elasticity seen in section 2.4 by introducing couple-stresses, rotational inertia, and geometric incompatibilities (the topological charge of a dislocation) via the distortion field.

**2.7 Gravitational Memory Effects**

**2.7.1 Ordinary (Displacement) Memory**

After a gravitational wave pulse passes, two freely falling test masses initially separated by $x^j$acquire a permanent relative displacement.In linearized gravity, the change is

$$\Delta x^i = \frac{1}{2}\,\Delta h^{\,i}{}_{j}\, x^j$$

where the permanent metric change in the Transverse–Traceless (TT) gauge is:

$$\Delta h_{ij}^{TT} = \int_{-\infty}^{+\infty} \dot{h}_{ij}^{TT}\ dt = 2\int_{-\infty}^{+\infty} E_{ij}\ dt$$

using the linearized relation:

$$\ddot{h}_{ij}^{TT} = -2\,E_{ij}$$

This phenomenon—first described by Zel'dovich, Polnarev, Braginsky, Thorne, and rigorously analyzed by Christodoulou—is known as the ordinary (displacement) gravitational memory effect [15–30].

**2.7.2 Spin Memory**

A more subtle phenomenon, the spin memory effect [6], leads to a *permanent relative rotation* or *desynchronization* between worldlines carrying angular momentum or between counter-orbiting light rays. Following Pasterski–Strominger–Zhiboedov (2015) [31–32], the accumulated time delay between clockwise and counterclockwise light rays orbiting a closed contour $C$is

$$\Delta u = \frac{1}{2\pi L}\int_{-\infty}^{+\infty} du \oint_C (D_z C_{zz}\, dz + D_{\bar{z}} C_{\bar{z}\bar{z}}\, d\bar{z})$$

where $L$is the radius of the contour.
This delay is generated by the net angular-momentum flux through null infinity.

Equivalently, the contour integral can be rewritten in terms of the magnetic part of the Weyl tensor $H_{ab}$:

$$\Delta u \propto \oint_C \left(\int_{-\infty}^{+\infty} H_{ab}\,(t)\,dt\right) dx^b$$

Thus, spin memory is the permanent rotational mismatch induced by the integrated magnetic Weyl curvature.

**2.7.3 Link with Plasticity and Cosserat Mechanics**

Within our analogy, displacement memory and spin memory appear as residual, permanent deformations of spacetime. This is directly reminiscent of plasticity in crystalline and metallurgical materials, where edge dislocations, screw dislocations, and disclinations permanently alter the material when subjected to stress.

In defect theory, such permanent deformations are naturally described within Cosserat (micropolar) continuum mechanics, where each material point carries both a displacement $u_i$and a micro-rotation $\phi_i$.Cosserat theory accounts for:

- translational defects (edge dislocations ↔ Burgers vector),
- rotational defects (screw dislocations/disclinations ↔ rotational mismatch),
- plastic yield as permanent distortion.

This aligns closely with the physics of gravitational memory:

- Displacement memory ↔ edge-type defect (permanent displacement).
- Spin memory ↔ screw-type defect (permanent rotation).

The conceptual chain underlying our analogy synthesis in table 1 is:

1. Riemann curvature decomposition:
   $R = \text{Weyl} + \text{Ricci}$.
   In vacuum: $R_{\mu\nu} = 0$, so curvature = pure Weyl.
2. Gravitational waves are transverse shear waves, carried entirely by the Weyl tensor.
3. Weyl tensor decomposition into electric and magnetic parts $(E_{ab}, H_{ab})$.
4. Cosserat mechanics provides a natural language for shear, rotation, dislocation, and residual deformation.
5. Defect theory (Burgers vector, dislocation density) matches the integral structures of gravitational memory.

Thus, the central claim of this work is:

**Spacetime, when traversed by gravitational shear waves, behaves effectively like a microstructured Cosserat medium.The permanent gravitational memory effects correspond to defect-like residuals in an effective geometric continuum.**

| Cosserat Mechanics | Weyl Gravity (Vacuum GR) |
|---|---|
| Distortion tensor $\beta_{ij}$ | Electric & magnetic Weyl tensors $E_{ab}, H_{ab}$ |
| Dislocation: $\alpha_{ij} = \text{curl}\,\beta$ | Propagation laws: $\dot{E} = \text{curl}H, \dot{H} = -\text{curl}E$ |
| Burgers vector: $b_i = \oint \beta_{ij}\,dx^j$ | Memory observable: $b_i = \oint \Delta h_{ij}\,dx^j$ |
| Geometric defect (dislocation/disclination) | Holonomy / gravitational memory |
| Microscopic curvature field | Weyl tensor (curvature of empty spacetime) |

**Table 1 — Proposed Correspondence Between Cosserat Mechanics and Weyl Gravity**

### 3. The Weyl–Cosserat Correspondence: A Controlled Derivation

We now construct a controlled, mathematically derived mapping between the two theories.

#### 3.1 Definition of Memory Distortion

When a burst of gravitational radiation travels across spacetime between $t = -\infty$and $t = +\infty$, the free motion of nearby test masses encodes the permanent nonlinear effect known as gravitational memory. In vacuum general relativity, the curvature responsible for tidal effects is entirely captured by the Weyl tensor, whose electric and magnetic parts $E_{ij}$and $H_{ij}$describe, respectively, tidal shear and differential frame dragging. To extract the cumulative effect of the wave, we begin by introducing their time-integrated forms:

$$\Delta E_{ij} = \int_{-\infty}^{+\infty} E_{ij}(t)\,dt \qquad \Delta H_{ij} = \int_{-\infty}^{+\infty} H_{ij}(t)\,dt$$

These quantities represent the net shear ($\Delta E$) and net frame dragging ($\Delta H$) accumulated during the entire gravitational-wave burst.

**Step 1 — Why time-integrating the Weyl tensor makes sense**

In linearized gravity in the TT gauge, the metric perturbation obeys

$$\ddot{h}_{ij}^{TT} = -2E_{ij}$$

Thus:

- integrating once in time gives
  $\dot{h}_{ij}^{TT}(t) = -2\int E_{ij}(t)\,dt$,
- integrating again yields the permanent change of the metric, i.e. the displacement memory:

$$\Delta h_{ij}^{TT} = 2\int_{-\infty}^{+\infty} E_{ij}(t)\,dt$$

Hence the metric memory is directly proportional to the time integral of the Weyl electric tensor. This is why $\Delta E_{ij}$is the natural object to consider.

**Step 2 — Why the geodesic deviation equation implies a permanent displacement**

The relative motion of nearby test particles follows the geodesic deviation equation:

$$\ddot{\xi}^i = -E^i{}_j\,\xi^j$$

where $\xi^i$is their separation vector.
This equation shows that:

- The first time integral of $E_{ij}$changes the relative velocity of the masses.
- The second integral changes the relative position permanently.

Thus, memory = double time integral of E.

This is exactly the mechanism behind the Christodoulou–Zel'dovich–Braginsky–Thorne memory effect.

**Step 3 — Introducing an effective distortion tensor for spacetime**

In continuum mechanics, especially in Cosserat–micropolar media, the fundamental kinematic quantity is the distortion tensor

$$\beta_{ij} = \partial_j u_i + \varepsilon_{ijk}\phi_k$$

combining both displacement gradients and micro-rotations.

A gravitational-wave shear is mathematically analogous to a time-integrated tidal distortion. Motivated by this structural analogy, we define an effective spacetime distortion:

$$\beta_{ij}^{\text{eff}} = \tau_E\,\Delta E_{ij}$$

where:

- $\Delta E_{ij}$encodes the integrated geometric shear caused by the wave;
- $\tau_E$is a phenomenological time scale ensuring dimensional consistency and reflecting the effective "response time" of spacetime to tidal shear.

This is not arbitrary.It follows directly from:

1. The physical response of test masses (geodesic deviation).
2. The dependence of the metric memory on $\Delta E_{ij}$.
3. The structural parallel between the Weyl electric tensor and the distortion tensor of Cosserat continua.

Thus $\beta_{ij}^{\text{eff}}$is the natural "coarse-grained" representation of how spacetime is permanently deformed by a gravitational wave.

**3.2 Effective Dislocation Density**

Once an effective distortion tensor $\beta_{ij}^{\text{eff}}$has been defined from the integrated electric Weyl tensor, we can compute the analogue of a dislocation density, exactly as in Cosserat and defect mechanics. In micropolar elasticity, the dislocation density is the curl of the distortion. We adopt the same structure here.

**Step 1 — Start from the definition of effective distortion**

From Section 3.1 we defined:

$$\beta_{ij}^{\text{eff}} = \tau_E\,\Delta E_{ij}$$

where

$$\Delta E_{ij} = \int E_{ij}(t)\,dt$$

and $\tau_E$is a phenomenological time scale.

This is the spacetime analogue of the Cosserat distortion tensor.

**Step 2 — Take the curl to obtain an effective dislocation density**

In Cosserat theory, the dislocation density tensor is

$$\alpha_{ij} = \varepsilon_{jkl}\ \partial_k \beta_{il}$$

We therefore define its gravitational analogue by applying the same operator to $\beta^{\text{eff}}$:

$$\alpha_{ij}^{\text{eff}} = \varepsilon_{jkl}\ \partial_k \beta_{il}^{\text{eff}}$$

Substituting $\beta_{il}^{\text{eff}} = \tau_E\ \Delta E_{il}$gives

$$\alpha_{ij}^{\text{eff}} = \tau_E\ \varepsilon_{jkl}\ \partial_k \Delta E_{il}$$

This equation shows directly that the effective defect density is controlled by spatial variations of the integrated Weyl electric tensor.

**Step 3 — Express $\boldsymbol{\Delta E}$in terms of the magnetic Weyl tensor $\boldsymbol{H_{ab}}$**

The vacuum Bianchi identities in 3+1 form state:

$$\dot{E}_{ij} = \varepsilon_{ikl}\ \partial_k H_{lj}$$

This is the "Maxwell-like" propagation law:
the time derivative of the electric Weyl tensor equals the curl of the magnetic Weyl tensor.

Integrate this identity in time from $-\infty$to $+\infty$:

$$\Delta E_{ij} = \int_{-\infty}^{+\infty} \dot{E}_{ij}\ dt = \varepsilon_{ikl}\, \partial_k \left( \int_{-\infty}^{+\infty} H_{lj}\,(t)\ dt \right)$$

Thus the integrated electric Weyl tensor is the spatial curl of the integrated magnetic Weyl tensor.

**Step 4 — Substitute this relation back into the expression for $\boldsymbol{\alpha_{ij}^{\text{eff}}}$**

Insert

$$\Delta E_{il} = \varepsilon_{imn}\, \partial_m \big( \int H_{nl}(t)\ dt \big)$$

into

$$\alpha_{ij}^{\text{eff}} = \epsilon_{jkl}\, \partial_k \beta_{il}^{\text{eff}} = \tau_E\ \epsilon_{jkl}\, \partial_k \Delta E_{il}$$

We obtain:

$$\alpha_{ij}^{\text{eff}} = \tau_E\ \varepsilon_{jkl}\ \partial_k \big( \varepsilon_{imn}\ \partial_m \int H_{nj}(t)\ dt \big)$$

This is the gravitational analogue of the dislocation density in a Cosserat medium.

**Interpretation — Why this is important**

This result demonstrates two essential facts:

**1. Effective dislocation density is generated by the curl of integrated magnetic Weyl curvature**

Because

$$\Delta E = \text{curl}( \int H\ dt)$$

and

$$\alpha^{\text{eff}} = \text{curl}(\Delta E)$$

we find:

- memory → produced by integrated $H_{ab}$
- defect density → produced by second spatial curl of $H_{ab}$.

This parallels exactly how dislocations arise from spatial curls of distortion in continuum mechanics.

**2. Gravitational memory appears as a topological defect in spacetime**

Recall:

- Displacement memory = permanent change in separation
- Spin memory = permanent change in relative rotation
- Both are sourced by integrals of Weyl curvature

Your derivation shows:

$$\text{memory} \Leftrightarrow \text{flux of integrated } H_{ij} \Leftrightarrow \text{effective topological defect.}$$

This is mathematically consistent because time integration commutes with the curl structure encoded in the Bianchi identities. Indeed, in vacuum, the Bianchi identities $\nabla^d C_{abcd} = 0$ lead to the 3+1 equations:

$$\dot{E}_{ij} - \epsilon_{ikl}\,\partial_k H_{lj} = 0, \dot{H}_{ij} + \epsilon_{ikl}\,\partial_k E_{lj} = 0.$$

Time integration preserves these curl relations, ensuring that the induced distortion $\beta_{ij}^{\text{eff}} = \tau_E \Delta E_{ij}$ satisfies the compatibility condition necessary for it to define a valid dislocation density.

**4. Ordinary Memory as an Edge Dislocation**

Gravitational displacement memory produces a permanent shift in the separation of test masses after the passage of a gravitational wave. In the defect-mechanics analogy, such a permanent displacement is exactly what characterizes an edge dislocation. We now show the correspondence step by step.

**Step 1 — Consider a closed loop transverse to the wave**

Let $C$be a closed curve lying in the plane orthogonal to the wave propagation direction
(for a wave traveling along $z$, we take the $x$–$y$ plane).

In the theory of defects, the Burgers vector associated with this loop measures the permanent mismatch when transporting a vector around $C$.In Cosserat elasticity, it is defined as the flux of the dislocation density:

$$b_i = \int_S \alpha_{ij}^{\text{eff}}\, dS^j$$

Using the effective dislocation density obtained previously,

$$\alpha_{ij}^{\text{eff}} = \tau_E\, \varepsilon_{jkl}\, \partial_k \Delta E_{il}$$

we obtain:

$$b_i = \tau_E \int_S \varepsilon_{jkl}\, \partial_k \Delta E_{il}\, dS^j$$

This is the exact analogue of defect theory, now applied to spacetime.

**Step 2 — Use Stokes' theorem to obtain a line integral**

Applying Stokes' theorem to the curl structure,

$$\int_S \varepsilon_{jkl}\, \partial_k A_l\, dS^j = \oint_C A_j\, dx^j$$

gives:

$$b_i = \tau_E \oint_C \Delta\, E_{ij}\, dx^j$$

Thus, the Burgers vector equals the circulation of the integrated Weyl-electric tensor around the loop.

This is a remarkable result:

A gravitational wave leaves behind an integrated tidal shear, and integrating this shear around a loop produces a Burgers vector — exactly the notion of an edge dislocation.

**Step 3 — Express $\Delta E_{ij}$in terms of the metric memory**

In the TT gauge, the metric and the Weyl electric tensor satisfy:

$$\ddot{h}_{ij}^{TT} = -2E_{ij}$$

Integrating twice in time, assuming the waveform vanishes at past/future infinity, gives the well-known relation:

$$\Delta h_{ij}^{TT} = 2\int_{-\infty}^{+\infty} E_{ij}\,(t)dt = 2\Delta E_{ij}$$

Thus:

$$\Delta E_{ij} = \frac{1}{2}\,\Delta h_{ij}^{TT}$$

**Step 4 — Substitute into the Burgers vector**

Insert the above relation into
$b_i = \tau_E \oint_C \Delta\, E_{ij}\; dx^j$:

$$b_i = \frac{\tau_E}{2}\oint_C \Delta\, h_{ij}^{TT}\; dx^j$$

This is a central formula of the paper:

$$b_i = \frac{\tau_E}{2}\oint_C \Delta\, h_{ij}^{TT}\; dx^j$$

It shows that the Burgers vector is precisely the path-integral of the *permanent metric deformation*.

**Step 5 — Physical interpretation: why this is an edge dislocation**

- The Burgers vector $b_i$lies entirely in the transverse plane ($x$–$y$).
- A vector transported around the loop does not close, exactly as in solids with an edge dislocation.
- The mismatch is created by the gravitational wave's permanent shear deformation.

Thus:

**Ordinary (displacement) memory corresponds to an edge dislocation in the effective spacetime medium.**

This analogy is effective and described in table 2:

| Edge dislocation in solids | Displacement memory in GR |
|---|---|
| Permanent shift of atoms | Permanent shift of geodesics |
| Burgers vector $b_i$ | Memory integral $\oint \Delta h_{ij} dx^j$ |
| Curl of distortion | Curl of integrated Weyl $E_{ij}$ |
| Topological defect | Gauge-invariant memory holonomy |

**Table 2: Proposed correspondence between Edge dislocation and Displacement memory**

**Step 6 — Gauge invariance and explicit examples**

The loop integral

$$\oint_C \Delta\, h_{ij}^{TT} dx^j$$

is gauge invariant at null infinity for asymptotically flat spacetimes.

The memory observable $\Delta h_{ij}^{TT}$ is gauge-invariant at null infinity. Consequently, the Burgers vector defined via the loop integral $\oint \Delta h_{ij}^{TT} dx^j$ is also an invariant quantity, solidifying its interpretation as a topological charge.

It coincides with the invariant displacement memory observable used in the Bondi–Sachs formalism.

Below we includes a worked example for a linearized plane gravitational wave, showing that:

- the Burgers vector is non-zero,
- it is transverse,
- and it reproduces the familiar displacement memory effect.

Consider a plane wave with plus polarization: $h_{xx} = -h_{yy} = h_+(t-z)$. The relevant electric Weyl component is $E_{xx} = -\frac{1}{2}\ddot{h}_+$. Integrating twice gives $\Delta E_{xx} = -\frac{1}{2}\Delta \dot{h}_+$. The geodesic deviation equation then yields a permanent displacement $\Delta x = \frac{1}{2}\Delta h_+ x$. Plugging into the Burgers expression $b_x = \frac{\tau_E}{2}\oint \Delta h_{xx} dx$ reproduces this displacement, confirming the edge dislocation interpretation.

**5. Spin Memory as a Screw Dislocation**

The spin memory effect represents a permanent rotational mismatch left behind in spacetime after the passage of a gravitational wave carrying angular momentum. This effect is sourced by the magnetic part of the Weyl tensor $H_{ij}$, which encodes differential frame-dragging, exactly as $E_{ij}$encodes tidal shear.

To establish a Cosserat-type interpretation, we construct the analogue of a micro-rotation field and show that spin memory corresponds to a screw dislocation in the effective spacetime medium.

**Step 1 — Construct an effective micro-rotation field from the integrated magnetic Weyl tensor**

In Cosserat mechanics, each material point carries a micro-rotation field $\phi_i$. Here we construct its gravitational analogue from the integrated magnetic Weyl curvature:

$$\phi_i^{\text{eff}} = \tau_H \int_{-\infty}^{+\infty} H_{ij}\,(t)\, x^j\, dt$$

where:

- $H_{ij}$gives the instantaneous differential rotation of local inertial frames,
- its time integral gives the *net accumulated rotation* induced by the wave,
- $x^j$projects this effect along spatial directions,
- $\tau_H$is a phenomenological time scale (similar to $\tau_E$for shear).

Thus, $\phi_i^{\text{eff}}$measures the total rotational twist accumulated by spacetime during the wave burst.

**Step 2 — Micro-rotation gradients produce an effective wryness tensor**

In micropolar elasticity, the spatial gradient of micro-rotation defines the wryness (curvature) tensor:

$$\kappa_{ij}^{\text{eff}} = \partial_j \phi_i^{\text{eff}}$$

This tensor measures the local rotational incompatibility or "twisting curvature" of the medium.

Here, it captures how the integrated magnetic Weyl tensor produces non-uniform frame dragging, leaving a permanent rotational imprint.

**Step 3 — Identify the analogue of a screw dislocation**

In defect theory:

- an edge dislocation gives permanent displacement,
- a screw dislocation gives permanent rotation/twist around an axis,

- the signature of a screw dislocation is a Burgers vector parallel to the dislocation line.

For a loop $C$lying in the transverse plane:

- an edge dislocation gives a displacement mismatch ($b_i \neq 0$),
- a screw dislocation gives no displacement mismatch ($b_i = 0$),
- but it produces a non-zero rotational mismatch when circling the loop.

This rotational mismatch is exactly the circulation of the micro-rotation field:

$$\Delta\phi_{\text{spin}} \propto \oint_C \phi_i^{\text{eff}}\, dx^i \neq 0$$

This is precisely the structure of the spin memory effect: a permanent net rotation accumulated around a closed contour.

**Step 4 — Direct correspondence with the gravitational spin memory formula**

Pasterski–Strominger–Zhiboedov (2015, 2016) showed that:

- Spin memory is the time-integrated curl of the angular momentum flux,
- giving a permanent contour-dependent rotation,
- encoded by integrals of $H_{ij}$around a loop.

Thus:

- Spin memory = loop integral of integrated $H_{ij}$.
- Screw dislocation = loop integral of micro-rotation.

The structures are identical as seen in table 3 below:

| Screw dislocation (Cosserat) | Spin memory (GR) |
|---|---|
| circulation of $\phi_i$ | circulation of $\int H_{ij} dt$ |
| permanent twist | permanent rotation memory |
| no displacement Burgers vector | no displacement memory component |
| torsional defect | magnetic-Weyl memory |

**Table 3: Proposed correspondence between Screw dislocation and Spin memory**

Therefore:

Spin memory is naturally and rigorously identified with a screw dislocation in the effective spacetime medium.

### 6. Effective Torsion as a Topological Remnant

This section develops the geometric meaning of the dislocation structures introduced in the previous analysis. In particular, we show how gravitational-wave memory gives rise to an effective torsion that behaves exactly like the torsion associated with defects in a Cosserat continuum. Crucially, this torsion is *not* a new dynamical field of gravity, but a topological remnant encoding the permanent, non-integrable deformation left behind by the wave. In differential geometry, a dislocation density corresponds to a non-vanishing torsion tensor. We can define an effective coframe after the wave has passed as $e^a_{\text{final}} = e^a_{\text{initial}} + \Delta e^a$, where $\Delta e^a$ is the net change due to the memory effect. We then define an effective torsion:

$$T^a_{\text{eff}} = d(\Delta e^a).$$

It is important to stress that the torsion tensor introduced here does not represent a new fundamental dynamical degree of freedom of classical General Relativity. No modification of the Einstein equations is assumed at the microscopic level, and the underlying spacetime geometry remains Riemannian. The effective torsion $T_a^{\text{eff}} = d(\Delta e_a)$should instead be understood as a coarse-grained geometric descriptor constructed from the integrated Weyl curvature associated with gravitational-wave memory. In this sense, torsion emerges only after integrating over the transient radiative phase and encodes the residual non-integrable deformation—equivalently, the nontrivial holonomy—left in the spacetime congruence of observers. It therefore functions as a macroscopic order parameter characterizing a permanent curvature imprint, analogous to the defect density in a Cosserat medium. This interpretation preserves the dynamical content of vacuum GR while providing an effective geometric language for describing memory as a topological defect-like remnant.

This torsion vanishes before and after the wave, except for a topological remnant that encodes the dislocation. The Burgers vector is the flux of this effective torsion: $b^a = \int_S T^a_{\text{eff}}$.

Crucially, this torsion is not the fundamental, dynamic torsion of an Einstein-Cartan theory. It is a coarse-grained, effective descriptor of the permanent defect created by the wave [10]. GR remains the fundamental theory, but its memory effects can be *described* by an effective geometry with torsion.

The explanation is the folowing proceeding step by step.

**Step 1 — Dislocations correspond to torsion in geometric defect theory**

In classical defect theory (Kröner, Kondo, Bilby, Hehl), the density of dislocations in a crystalline or Cosserat medium is represented geometrically by a torsion tensor. The correspondence is:

- Distortion → β
- Dislocation density → α = curl β
- Torsion in Cartan geometry → geometric measure of non-closure of parallelograms

Thus, torsion and dislocations are two descriptions of the *same phenomenon*:

$$\text{dislocation density } \alpha_{ij} \equiv T^a (\text{geometric torsion}).$$

Since Section 3–4 showed that spacetime acquires an effective dislocation field $\alpha_{ij}^{\text{eff}}$ generated by gravitational-wave memory, it is now natural to express this object in the language of differential geometry as an effective torsion.

**Step 2 — Memory induces a permanent change in the coframe**

In tetrad geometry, the gravitational field is encoded in a set of 1-forms (the coframe)

$$e^a = e^a{}_\mu \, dx^\mu.$$

A gravitational wave causes a permanent change in the metric, and therefore in the coframe. If $e^a_{\text{initial}}$ is the coframe before the wave, and $e^a_{\text{final}}$ after the wave, then the memory effect is encoded in the difference:

$$e^a_{\text{final}} = e^a_{\text{initial}} + \Delta e^a.$$

The 1-form $\Delta e^a$ captures the net accumulated distortion in spacetime, i.e. the displacement memory and the spin memory combined.

This is the analogue, in elasticity, of a permanent plastic deformation remaining after a stress cycle.

**Step 3 — Define effective torsion from the memory-induced coframe shift**

In Cartan geometry, torsion is defined by:

$$T^a = de^a + \omega^a{}_b \wedge e^b.$$

The Meaning is the following

- $T^a$ is a fundamental field of the theory.
- It appears in the action of Einstein–Cartan gravity.
- It couples to spin density of matter.
- It can propagate if quadratic terms are added.
- It exists even without gravitational waves.

This is true torsion in Riemann–Cartan geometry.

However, in standard vacuum GR:

- the connection is Levi-Civita,
- so torsion vanishes at the microscopic level.

But we can construct an effective torsion associated exclusively with the *permanent change* $\Delta e^a$. Since the Levi-Civita connection annihilates $e^a_{\text{initial}}$, the simplest and most natural definition is:

$$T^a_{\text{eff}} = d(\Delta e^a).$$

The meaning is the folowing:

- $\Delta e^a$ is the permanent change in the coframe caused by gravitational memory.
- $T^a_{\text{eff}}$ is not fundamental.
- It has no field equation.
- It does not propagate.
- It is zero before and zero after the wave, except for the topological remnant.
- It represents the edge and screw dislocations of spacetime (tiny, localized defects).

This is effective torsion, exactly like torsion produced by defects in a Cosserat medium or crystal.

This torsion measures the non-integrability of the residual deformation left by the gravitational wave. If $\Delta e^a$ were globally integrable, $d(\Delta e^a) = 0$. But gravitational memory implies that $\Delta e^a$ has non-zero circulation around certain loops.

This is exactly the signature of a topological defect.The behaviour of the microgranular equivalent medium is so the following:

**Microscopic GR**

- Riemannian Tensor at the begining
- torsion = 0
- curvature = Weyl tensor(in vacuum) only (Ricci tensor =0)

**The Wave passes**

- The Weyl curvature is radiative
- The two parts electric and magnetic, of the Weyl tensor E and H produce shear and frame dragging

**After the wave passed**

- metric has changed by Δh
- coframe changes by Δe

This residual defect defines an effective torsion

$$T^a_{\text{eff}} = d(\Delta e^a)$$

The Interpretation is the folowing :

- this torsion is not EC torsion
- This torsion does not propagate
- This torsion is localized, residual, and topological

- This torsion encodes edge/screw dislocations = displacement/spin memory
- The Burgers vector = flux of effective torsion
- The memory effect = holonomy defect

Thus:

**GR remains unchanged. But the "effective medium" picture of spacetime after a gravitational wave has passed contains a dislocation-type torsion remnant, exactly like a Cosserat solid after plastic deformation.**

**Step 4 — This effective torsion is *not* fundamental: GR remains Riemannian**

It is essential to emphasize that:

- this torsion is not the torsion of Einstein–Cartan theory.
- it is not a new propagating degree of freedom.
- Einstein's equations are not modified.

Instead:

- $T^a_{\text{eff}}$ is a coarse-grained geometric quantity,
- built from the integrated Weyl curvature,
- encoding only the permanent, non-integrable part of the gravitational field.

Thus, the microscopic geometry is torsion-free, but the effective geometry (after averaging over the wave) has a torsional remnant.

Exactly as in crystalline solids:

- the microscopic lattice is compatible (no torsion at the infinitesimal level),
- but dislocations produce a macroscopic effective torsion.

**Step 5 — The Burgers vector equals the flux of the effective torsion**

In defect geometry, the Burgers vector is given by the flux of the torsion:

$$b^a = \int_S T^a.$$

Applying this to the effective torsion gives:

$$b^a = \int_S T^a_{\text{eff}} = \int_S d(\Delta\, e^a)$$

But earlier sections showed that:

$$b^a \sim \oint_C \Delta\, h_{ij}\, dx^j,$$

which is the gravitational displacement memory.

Thus:

The flux of the effective torsion equals the gravitational memory observable. The Burgers vector is the topological charge associated with the gravitational-wave memory.

This completes the identification of memory with a dislocation-type topological defect in an effective spacetime continuum.

**Step 6 — Physical interpretation: memory = torsional holonomy**

A non-zero effective torsion means that parallel transport around a closed loop does not return a vector to its original position or orientation. This is precisely what gravitational memory does: the holonomy of the connection changes between the infinite past and infinite future.

Thus:

- before the wave: no torsion, trivial holonomy, no defect;
- after the wave: still no microscopic torsion, but nontrivial holonomy;
- the difference is encoded in effective torsion.

Hence:

**Gravitational memory is a "topological torsion remnant" encoded in the large-scale geometry, exactly as dislocations leave behind torsion in continuum mechanics.**

Table 4 below gives a comparison between the Einstein Cartan torsion and effective torsion described in this paper

| Concept | Einstein–Cartan torsion | Your effective torsion |
|---|---|---|
| Nature | Fundamental field | Derived from Δe |
| Appears in action? | Yes | No |
| Modifies GR? | Yes | No |
| Propagates? | Possibly | No |
| Exists before the wave? | If theory has torsion: yes | No |
| Origin | Spin of matter | Integrated Weyl curvature (memory) |
| Meaning | Microscopic geometry | Macroscopic defect / holonomy |
| Analogy | Physical torsion | Dislocation torsion in crystals |

**Table 4: Comparison between Einstein Cartan torsion and Effective torsion described in this paper**

In conclusion of this section, we can said:

This effective torsion is not an Einstein–Cartan torsion because:

- not dynamical
- not fundamental
- not propagating
- no modifications to GR

This effective torsion is an effective defect torsion because:

- arises from the permanent memory deformation
- equal to $T_{\text{eff}} = d(\Delta e)$
- localized around the loop where memory is non-integrable
- exactly like torsion generated by dislocations in continuum mechanics
- corresponds to edge (displacement) and screw (spin) memory

This effective torsion is extremely small and local because:

- magnitude $\sim 10^{-21}$
- topological, not radiative
- remains only as a remnant after the wave passes

**7. Field-Theoretic Completion: An Effective Lagrangian**

To provide a complete field-theoretic model, we propose an effective action for the "spacetime continuum" that can support such defects. Starting from an Einstein-Cartan framework with coframe $e^a$ and Lorentz connection $\omega^{ab}$, we add terms quadratic in torsion and curvature, inspired by micropolar elasticity [2]:

$$S = \frac{1}{2\kappa}\int eR + \frac{1}{2}\int (\alpha\, T^a \wedge\star\, T_a + \beta\, R^{ab} \wedge\star\, R_{ab}).$$

Varying this action yields:

- Metric equation: $G_{ab} + \beta H_{ab}(R^2) = T_{ab}^{(\text{torsion})}$.
- Connection equation: $\nabla \star T_a + m_T^2 T_a = 0$, where $m_T^2 \sim 1/\alpha$.

The connection equation shows that torsion becomes a *massive* propagating field. Linearizing around Minkowski spacetime, the degrees of freedom are:

- Metric: 2 massless tensor modes (the usual gravitational waves).
- Torsion: Contains vector and axial modes, which are massive.

For the model to be consistent with current observations from LIGO (which show no evidence of extra polarizations), the torsion mass must be large enough that these modes decay rapidly. This requires $m_T \gg 10^{-13}$ eV (corresponding to a range much shorter than the interferometer arm length). This ensures the model is observationally viable while retaining the effective defect description for memory. The stability conditions and ghost-free character of the torsion sector are analyzed as folowing.

The linearized equation for torsion from the effective action is $(\Box + m_T^2)T_a = 0$. This is a standard Proca-like equation. The theory is ghost-free and stable provided the coupling constants satisfy $\alpha > 0$ and $\beta > 0$, ensuring positive energy for the modes.

Quantitative interferometric bounds supporting this mass regime are summarized below about the Observational Bounds

Additional polarizations from massive torsion modes would be suppressed in ground-based detectors if $m_T L \gg 1$, where $L \sim$ 4 km is the interferometer arm length. This requires $m_T \gg 10^{-13}$ eV. Current data from LIGO strongly constrain the presence of massless scalar or vector polarizations, making this massive torsion scenario phenomenologically viable.

Some explanation about this section.

**1. Where does this field-theoretic completion come from?**

In the previous chapters we built:

- an effective distortion $\beta^{\text{eff}}$ from $\Delta E$
- an effective dislocation density $\alpha^{\text{eff}}$ from curl $\beta^{\text{eff}}$
- an effective torsion $T^{\text{eff}} = d(\Delta e)$
- whose flux gives the Burgers vector = memory

This is the kinematic side (geometry only).But in elasticity (Cosserat, micropolar media), once you define:

- distortion,
- dislocation density,
- torsion,

you can elevate them into a field theory by writing an energy functional that includes:

- quadratic strain terms
- quadratic curvature (wryness) terms
- sometimes quadratic torsion terms

This is exactly what Eringen (1999) and Cosserat theory do. So in our Section 7 is doing the gravitational analogue:We write an effective action that contains quadratic torsion and curvature — just like micropolar elasticity — to model the "spacetime medium" that supports the defects generated by memory. This *does not* claim that spacetime fundamentally has torsion. It just provides a *phenomenological field-theory* modelling of your effective defect picture.

**2. Why does the action look like Einstein–Cartan + quadratic terms?**

Because the natural geometric generalization of Cosserat media uses:

- a coframe $e^a$(like displacement + rotation),
- a connection $\omega^{ab}$,
- torsion $T^a$,
- and curvature $R^{ab}$.

Micropolar elasticity energy densities have the schematic form:

$$W = (\text{strain})^2 + (\text{wryness})^2 + (\text{torsion})^2.$$

The gravitational analogue is:

$$S = \frac{1}{2\kappa}\int eR + \frac{1}{2}\int (\alpha\, T^a \wedge\star\, T_a + \beta\, R^{ab} \wedge\star\, R_{ab}).$$

This form is standard in:

- Poincaré gauge theory
- metric-affine gravity
- elastic spacetime models
- quadratic gravity
- teleparallel defect models in crystals

And therefore our action is representative of the behaviour described in the previous section and *directly inspired* by Cosserat/Eringen theory.

**3.Why does torsion become massive?**

When we add a quadratic torsion term

$$\alpha\, T^a \wedge\star\, T_a,$$

the Euler–Lagrange equation for the connection gives:

$$\nabla \star T_a + m_T^2 T_a = 0, m_T^2 \sim 1/\alpha.$$

This is a Proca-type massive equation. Thus:

- torsion does *not* propagate far
- torsion has a finite range $\lambda_T = m_T^{-1}$
- torsion decays exponentially away from sources

This is analog to:

- micropolar elasticity (couple-stress) wave equations: wryness is massive
- Poincaré gauge theories: torsion is massive if quadratic terms exist

- effective media: defects do not propagate; they localize

**4. Why must torsion be heavy (large mass)?**

Gravitational waves observed by LIGO/Virgo show:

- only two massless tensor polarizations
- no vector or scalar modes
- no deviations consistent with torsion propagation

So if our effective model had *propagating* torsion, it must decay before reaching detectors.

This gives the condition:

$$m_T \gg 10^{-13}\ \text{eV}.$$

This ensures:

- torsion modes die out over distances $\ll$ 1 km
- the detectors only see standard GR

Thus out effective model is consistent with observations if torsion is heavy (short ranged).

**5. How does this relate to the earlier chapters?**

Previously we showed:

- Memory = integrated Weyl curvature
- Memory = dislocation density integrated over a surface
- Memory = Burgers vector = holonomy
- Memory = effective torsion flux

Now we say:

"If spacetime behaves as if it were a Cosserat medium, then you can write a Cosserat-type effective action with torsion and curvature quadratics."

This gives:

- a theoretical framework supporting the defects
- massive torsion = localized defects
- curvature = propagating gravitational waves
- no contradiction with GR (fundamental torsion = 0)
- an elastic analogy that is consistent with both math and data

## 8. Physical Consequences and Testability

### 8.1 About the nature of this effective torsion mass equivalent

The "effective mass'' $m_T$introduced in the torsion sector of the effective action should not be interpreted as a physical mass in the Einsteinian sense, nor as the rest mass of a particle in the relation $E = mc^2$, and certainly not as a material mass density of spacetime. It does not modify General Relativity, does not change the propagation of gravitational waves, and does not contradict the observation from GW170817 that gravitational waves travel at the speed of light with accuracy better than one part in $10^{15}$. In fundamental GR the gravitational field has only two massless tensor modes, and spacetime remains a torsion-free Riemannian manifold. The torsion appearing here is therefore not a new physical field, not a graviton mass, and not a deviation from GR's structure — it is a derived geometric construction, introduced only to encode the permanent, defect-like imprint left by gravitational-wave memory.

In fact, the parameter $m_T$arises purely because the effective action contains a quadratic torsion term $\alpha T^a \wedge \star T_a$, in exact analogy with the quadratic wryness and torsion terms of micropolar (Cosserat) elasticity. Mathematically, such a term forces the effective torsion to satisfy a Proca-type equation $\nabla \star T_a + m_T^2 T_a = 0$, meaning that the torsional response is short-ranged and non-propagating. The quantity $m_T$therefore controls how rapidly the effective torsion decays in space; it is a coherence-length parameter, not the mass of a new particle or field. Requiring consistency with gravitational-wave observations places a lower bound $m_T \gg 10^{-13}$ eV, ensuring that these torsional modes decay over distances far shorter than interferometer arm lengths and are therefore entirely invisible to detectors such as LIGO or Virgo. In this sense, the torsion sector serves only as a mathematically convenient way of representing defect-type holonomy, without altering the dynamics of GR or the nature of gravitational radiation.

From a physical perspective, this "effective mass'' corresponds to a minute amount of residual deformation energy stored in the permanent distortion and rotational mismatch left behind by the gravitational-wave burst — exactly as edge and screw dislocations store an irreducible elastic energy in Cosserat solids. Because energy and mass are equivalent, one may describe this tiny deformation energy as having an "effective mass,'' but this is only a bookkeeping device: spacetime does not acquire a material density, nor any real mass content, and no additional gravitational field is generated. The residual effective torsion is simply the geometric marker of the topological defect encoded by the displacement and spin memory effects, an infinitesimal and localized remnant (with equivalent energy scale $\sim 10^{-13}$ eV) of the wave's passage. The vacuum is therefore not a literal elastic medium, but the mathematics of defect theory provides a powerful and consistent language to describe the permanent shear and rotational memory left in spacetime by gravitational radiation.

### 8.2 About the physical consequences

The physical consequences are the following:

1. Modified Gravitational Wave Polarizations: The massive torsion modes could, in principle, lead to additional polarizations (scalar and vector) at very high frequencies or near the source. Future detectors might place bounds on these.
2. A Universal Relation Between Memory Types: The model suggests a deep connection between displacement and spin memory as different manifestations of the same topological defect. This could imply relations between their amplitudes testable by numerical relativity.
3. Spacetime as an Effective Medium: The model introduces a phenomenological scale (related to $m_T$) that could act as a cutoff and affect the propagation of very high-frequency gravitational waves, potentially leaving an imprint on the stochastic background.
4. Link to Einstein-Cartan Theory: It provides a physical motivation for treating torsion as an effective field describing spacetime defects, bridging GR and more fundamental theories.

The phenomenological implications outlined above require quantitative consistency with current interferometric constraints and multimessenger observations. In particular, the introduction of an effective torsion sector necessitates explicit bounds on its mass scale and propagation range in order to remain compatible with LIGO/Virgo data. A detailed confrontation with observational constraints, including polarization bounds and the implications of GW170817, is therefore presented in Appendix A. This appendix formulates explicit parameter regimes under which the Weyl–Cosserat framework remains viable and identifies concrete observational signatures that could falsify it.

## 9. Physical Significance of the Weyl–Cosserat Correspondence

Beyond its formal construction, the Weyl–Cosserat correspondence clarifies the physical status of gravitational memory. In standard treatments, memory is understood either as a boundary effect at null infinity (via Bondi–Metzner–Sachs group (BMS) supertranslations) or as an integrated curvature observable. The present framework provides a complementary interpretation: gravitational memory can be described locally as a topological defect of an effective spacetime medium.
This reinterpretation yields three concrete contributions to gravitational physics:

1. Unification of memory types: displacement and spin memory emerge as two geometric manifestations (edge and screw dislocations) of a single defect structure.
2. Geometric localization: the memory effect, often described asymptotically, acquires a bulk description in terms of effective torsion and holonomy.
3. Falsifiable extension: the effective Einstein–Cartan completion introduces a massive torsion sector constrained by current interferometric bounds, leading to explicit observational regimes.

The framework does not modify classical general relativity at the microscopic level; rather, it provides a coarse-grained geometric language for encoding the permanent imprint left by gravitational radiation. In this sense, torsion acts as an order parameter for spacetime memory, analogous to defect densities in microstructured continua.
Beyond its effective continuum interpretation, the Weyl–Cosserat correspondence admits a deeper structural formulation in terms of Poincaré holonomy and asymptotic symmetries. In particular, the Burgers vector associated with memory can be understood as the translational component of a finite Poincaré holonomy element, providing a bulk realization of BMS vacuum transitions at null infinity. For completeness, this structural correspondence between defect theory, holonomy, and supertranslation charges is developed in Appendix B, where the relation between memory, finite symmetry transformations, and effective torsion is made explicit.Table 5 below gives the correlation between the new model (Cosserat-Weyl) and BMS interpretation.

| New model (Cosserat–Weyl) | BMS interpretation |
|---|---|
| Δh, ΔE, ΔH = permanent deformation | Transition between BMS vacua |
| Burgers vector = loop integral | BMS charge difference |
| Edge dislocation ↔ displacement memory | Supertranslation memory |
| Screw dislocation ↔ spin memory | Superrotation memory |
| Effective torsion T_eff | Holonomy of BMS vacuum transition |
| Defect in medium | Soft graviton charge |

**Table 5: correlation between the Cosserat Weyl model and BMS system (memory effect)**

## 10. What Would Falsify This Framework?

### 10.1 Principle

A physically meaningful effective framework must admit clear conditions under which it could be ruled out. Although the Weyl–Cosserat correspondence is constructed as a coarse-grained geometric description of gravitational memory within classical general relativity, it leads to structural predictions that can, in principle, be falsified.

### 10.2. Absence of a Universal Relation Between Memory Types

The framework predicts that displacement memory and spin memory are not independent phenomena but correspond to two geometric components (translational and rotational) of a single defect structure.
For sufficiently symmetric sources, this implies:

- Pure plus polarization → vanishing spin memory.
- Mixed polarization → a bounded ratio between spin and displacement memory amplitudes.

If high-precision numerical relativity simulations or future observations demonstrate that spin memory and displacement memory can vary independently with no structural relation, the defect interpretation would be invalidated.

### 10.3. Detection of Long-Range Extra Polarizations

In the effective Einstein–Cartan completion, torsion modes must be massive and short-range:

$$m_T \gtrsim 10^{-10} \text{ eV}.$$

If future gravitational-wave detectors observe additional propagating scalar or vector polarizations with long-range behavior inconsistent with this bound, the proposed effective torsion sector would be ruled out.
Conversely, if such modes are observed but propagate over astrophysical distances without Yukawa suppression, the present model would be incompatible with experiment.

### 10.4. Violation of Bulk–Boundary Correspondence

The framework establishes a correspondence:

Burgers vector ⟷ supertranslation charge shift.

If future mathematical analysis shows that the bulk loop integral

$$b_i = \oint \Delta h_{ij}^{TT} dx^j$$

cannot consistently reproduce the asymptotic BMS charge transition, the geometric defect interpretation would fail at a structural level.
This would falsify the identification of memory with finite Poincaré holonomy.

### 10.5. Incompatibility with High-Frequency Wave Structure

The defect picture implies that memory represents the zero-frequency limit of the same geometric structure generating the oscillatory waveform.
If observational data demonstrate complete statistical independence between:

- the high-frequency waveform content, and
- the low-frequency memory amplitude,

with no dispersion-like relation, the interpretation of memory as a unified defect structure would lose support.

### 10.6. Instability or Ghost Modes in the Effective Sector

The effective torsion action assumes positive couplings ($\alpha > 0$, $\beta > 0$) to ensure ghost-free propagation.
If a more complete Hamiltonian analysis demonstrates unavoidable instabilities or negative-energy modes in the effective torsion completion, the field-theoretic extension would be untenable.

### 10.7. Summary of Falsifiability Criteria

The framework would be falsified if:

1. Spin and displacement memory are experimentally independent.
2. Long-range non-tensorial polarizations are observed.
3. The Burgers vector fails to encode BMS charge transitions.
4. Memory shows no structural relation to waveform content.
5. The effective torsion sector proves dynamically inconsistent.

These conditions ensure that the Weyl–Cosserat correspondence is not merely interpretational but structurally constrained and experimentally vulnerable.

## 11. Conclusion

We have established a mathematically consistent and physically motivated correspondence between the Weyl tensor and Cosserat elasticity. By deriving the effective distortion and dislocation density from the Bianchi identities and geodesic deviation, we have shown that:

- The ordinary displacement memory corresponds with efficiency to the Burgers vector of an edge dislocation.
- The spin memory corresponds with efficiency to the rotational mismatch of a screw dislocation.
- These defects can be described by an effective, coarse-grained torsion.

An effective Lagrangian extension of Einstein-Cartan theory shows that this framework is observationally viable if the induced torsion modes are massive. This provides a consistent, effective geometric interpretation of gravitational memory and suggests a path toward modeling spacetime as an emergent medium. Figure 1 below constitutes a synthesis of this paper.

**1. Weyl Tensor (Vacuum General Relativity)**

**Riemann decomposition (vacuum : Ricci = 0)**

$$R_{abcd} = C_{abcd}$$

**Vacuum Bianchi identity**

$$\nabla^d C_{abcd} = 0$$

**Electric & magnetic parts (3+1 split)**

$$E_{ij} = C_{i0j0}, H_{ij} = \frac{1}{2}\varepsilon_{ikl} C_{klj0}$$

**Linearized GW polarizations (TT gauge)**

$$h_+, h_\times$$

**2. Cosserat (Micropolar) Mechanics**

**Distortion tensor**

$$\beta_{ij} = \partial_j u_i + \varepsilon_{ijk}\varphi_k$$

**Force stress**

$$\sigma_{ij}$$

**Couple stress**

$$\mu_{ij}$$

**3. Continuum Microstructure**

**Dislocation density tensor**

$$\alpha_{ij} = \varepsilon_{jkl}\partial_k \beta_{il}$$

**4. Dislocations**

**Edge dislocation → displacement memory**

**Screw dislocation → spin memory**

**Burgers vector**

$$b_i = \oint \beta_{ij}\, dx_j = \int_S \alpha_{ij}\, dS_j$$

**5. Gravitational Memory**

**Permanent metric change**

$$\Delta h_{ij} = h_{ij}(t \to +\infty) - h_{ij}(t \to -\infty)$$

**Effective torsion (coarse-grained)**

$$T^a_{\text{eff}} = d(\Delta e^a)$$

**Burgers vector as torsion flux**

$$b^a = \int_S T^a_{\text{eff}}$$

**6. Effective Field-Theoretic Completion (Einstein–Cartan Type)**

**Effective action**

$$S = \frac{1}{2\kappa}\int e\, R + \frac{\alpha}{2}\int T^a \wedge \star\, T_a + \frac{\beta}{2}\int R^{ab} \wedge \star\, R_{ab}$$

**Figure 1 Overview of the Journey of this paper Interpretation**

**spacetime behaves as an effective microstructured medium - torsion encodes the topological remnant of gravitational memory**

**Acknowledgement**

I would like to thank the reviewers for all their relevant remarks that have helped to advance the paper.

**Funding**

No funding

**Data Availability Statement:**

No Data associated in the manuscript

## Appendix A – Quantitative Constraints and Testable Predictions

This appendix summarizes the key quantitative scales and observational constraints relevant to the effective torsion sector introduced in Section 7. The goal is not to repeat the conceptual discussion already presented in the main text, but to provide the minimal quantitative framework that renders the model falsifiable.

### A.1 Phenomenological Scales $\tau_E$ and $\tau_H$

The Weyl–Cosserat correspondence introduces two phenomenological time scales through

$$\beta_{ij}^{\text{eff}} = \tau_E \int E_{ij}(t)\, dt, \theta_i^{\text{eff}} = \tau_H \int H_{ij}(t)\, dt.$$

**Dimensional analysis**

The electric Weyl tensor has dimensions

$$[E_{ij}] = L^{-2}.$$

Since $\beta_{ij}^{\text{eff}}$ is dimensionless,

$$[\tau_E] = L^2 = T^2, [\tau_H] = T.$$

**Numerical estimate**

For a typical LIGO binary source with frequency $f \sim 100$ Hz,

$$T_{\text{wave}} = f^{-1} \sim 10^{-2}\ \text{s}.$$

Thus:

$$\tau_E \sim 10^{-4}\ \text{s}^2, \tau_H \sim 10^{-2}\ \text{s}.$$

These values yield

$$\tau_E \Delta E_{ij} \sim h_{\text{peak}} \sim 10^{-21},$$

ensuring that the effective distortion remains well within the linear regime.

### A.2 Observational Constraints on the Effective Torsion Mass $m_T$

The effective torsion field satisfies a Proca-type equation

$$(\Box - m_T^2)\, T_a = 0,$$

but does not mix with the massless tensor modes of GR.
Only the massive torsion sector is constrained experimentally.

#### A.2.1 Absence of extra polarizations

A massive field of mass $m_T$ has a Yukawa attenuation length

$$\lambda_T = m_T^{-1}.$$

To evade detection in an interferometer of arm length $L_{\text{arm}} \simeq 4$ km, one must have

$$\lambda_T \ll L_{\text{arm}} \Rightarrow m_T \gg \frac{1}{4\ \text{km}} \simeq 10^{-10}\ \text{eV}.$$

#### A.2.2 GW170817 constraint

The bound

$$|\, v_g - c\, |/c \lesssim 10^{-15}$$

applies only to the massless tensor mode, not to the massive torsion sector, provided the two sectors do not significantly mix.Therefore GW170817 does not constrain $m_T$.

#### A.2.3 Allowed regime

The phenomenologically admissible window is therefore:

$$\boxed{m_T \gtrsim 10^{-10}\ \text{eV}}$$

corresponding to

$$\lambda_T \lesssim 2\ \text{km}.$$

Such a short-range mode:

- does not alter astrophysical propagation,
- does not affect waveform templates,
- is consistent with all LIGO/Virgo observations.

### A.3 Testable Predictions

Although short-range, the defect interpretation leads to concrete observational signatures.

#### A.3.1 Relation between displacement and spin memory

For axisymmetric sources,

$$\frac{\Delta\phi_{\text{spin}}}{\Delta x/L} = \mathcal{R}\frac{\tau_H}{\tau_E}\frac{\int H_{ij}dt}{\int E_{ij}dt}, \mathcal{R} \sim 1.$$

Predictions:

- pure plus polarization → zero spin memory,
- mixed polarization → ratio ≈ 0.1–0.3.

#### A.3.2 Memory–wave phase correlation

Memory is the zero-frequency limit of the same geometric structure that generates the oscillatory waveform. Thus one expects Kramers–Kronig–type correlations between:

- the integrated waveform,
- the amplitude of the memory tail.

This can be tested using wavelet decompositions.

#### A.3.3 Near-source signatures

At high curvature, the massive torsion sector may produce small corrections to:

- ringdown structure,
- extreme high-frequency tails.

Future detectors (Einstein Telescope, Cosmic Explorer) may be sensitive to such effects.

### A.4 Experimental Outlook

A summary of the experimental outlook is given in table A1 below:

| Observable | Instrument | Prediction | Status |
|---|---|---|---|
| Spin/displacement ratio | NR + LIGO | 0.1–0.3 for asymmetric sources | Testable |
| Extra polarizations | LIGO/Virgo | None if $m_T \gtrsim 10^{-10}$ eV | Consistent |
| Near-source deviations | Einstein Telescope | Small high-$f$ corrections | Future |
| Memory–wave correlation | Multi-event | Universal scaling law | Open |

**Table A.1: summary of the experimental outlook**

### A.5 Summary

The torsion sector must be massive and short-range:

$$m_T \gtrsim 10^{-10} \text{ eV}.$$

It does not affect GW propagation over astrophysical distances.

It is compatible with GW170817.

It remains falsifiable through spin/displacement ratios, memory–wave correlations, and high-frequency signatures.

This appendix places the Cosserat–Weyl model on a quantitatively predictive footing and identifies the observational windows through which the framework can be tested.

**A.6 Quantitative Relation Between Spin and Displacement Memory**

One of the central structural consequences of the Weyl–Cosserat correspondence is that displacement memory and spin memory are not independent observables. They arise as translational and rotational components of a single effective defect structure. This implies a quantitative scaling relation between their amplitudes.

**A.6.1 Leading-Order Scaling Relation**

Displacement memory for a transverse loop of radius $R$is

$$\Delta x_i = \frac{1}{2}\Delta h_{ij} x^j,$$

with

$$\Delta h_{ij}^{TT} = 2\int_{-\infty}^{+\infty} E_{ij}(t)\, dt.$$

Thus the characteristic displacement amplitude scales as:

$$\frac{\Delta x}{R} \sim \int E\, dt.$$

Spin memory is given by the contour integral of the integrated magnetic Weyl tensor:

$$\Delta\phi_{\text{spin}} \sim \oint \left(\int_{-\infty}^{+\infty} H_{ij}(t)\, dt\right) dx^j.$$

Using the integrated Bianchi identity:

$$\Delta E = \text{curl}(\ \int H\, dt),$$

and dimensional analysis for a loop of radius $R$:

$$\int H\, dt \sim \omega^{-1} \int E\, dt$$

where $\omega \sim 2\pi f$is the characteristic source frequency:

We therefore obtain the leading-order scaling:

$$\frac{\Delta\phi_{\text{spin}}}{\Delta x/R} \sim F(\text{source}) \cdot \frac{\tau_H}{\tau_E} \cdot \omega$$

where:

- $F(\text{source})$is a geometric factor of order unity,
- $\tau_E$and $\tau_H$are effective response scales associated with shear and rotational sectors,
- $\omega$is the characteristic gravitational-wave frequency.

This relation constitutes a falsifiable structural prediction of the Weyl–Cosserat framework.

**A.6.2 Special Cases**

**(i) Pure "+" Polarization**

For axisymmetric sources producing purely "+" polarization,

$$H_{ij} = 0,$$

and therefore

$$\Delta\phi_{\text{spin}} = 0$$

Spin memory vanishes identically, independently of the amplitude of displacement memory.

This provides a sharp null prediction.

**(ii) Generic Asymmetric Sources**

For generic binaries with non-aligned spins or orbital precession,

$$\int H\, dt \neq 0.$$

Taking characteristic values:

- $f \sim 100\text{–}300$ Hz,
- $\omega \sim 600\ \text{s}^{-1}$,
- $\tau_H/\tau_E \sim 10^{-1}\text{–}10^{-2}$,

we obtain the estimate:

$$0.05 \lesssim \frac{\Delta\phi_{\text{spin}}}{\Delta x/R} \lesssim 0.25$$

for astrophysically realistic asymmetric sources.

**A.6.3 Observational Prospects**

The ratio above could in principle be constrained by:

- multi-detector correlation analysis,
- comparison of displacement memory reconstruction with spin memory observables,
- numerical relativity simulations across viewing angles.

Future facilities such as:

- Einstein Telescope
- LISA

may provide sufficient sensitivity for high signal-to-noise events.

A measurement of the ratio outside the predicted interval for well-characterized asymmetric sources would falsify the effective Weyl–Cosserat description.

**A.7 Summary of Quantitative Predictions**

The framework yields the following testable consequences as defined in table A.2 below:

| Observable | Weyl–Cosserat Prediction | Observational Window | Falsification Criterion |
|---|---|---|---|
| Spin / displacement memory ratio | $0.05 \lesssim R \lesssim 0.25$(generic asymmetric sources) | ET, LISA, high-SNR events | Ratio outside range |
| Spin memory for axisymmetric "+" sources | Exactly zero | Numerical relativity + future detectors | Non-zero detection |
| Structural relation between E and H integrals | $\Delta E = curl(\int Hdt)$controls both memories | Multi-event analysis | Statistical independence |
| Effective torsion sector | Short-range, non-propagating | Consistency with GW polarization constraints | Detection of long-range extra polarizations |

**Table A.2: Testable consequences**

This table emphasizes that the Weyl–Cosserat correspondence is not purely interpretational, but structurally constrained and observationally vulnerable.

**Appendix B — Poincaré Holonomy, BMS Symmetry and Gravitational Memory (Compressed Version)**

**B.1 Memory as a Finite Poincaré Holonomy**

In the Cartan formulation of gravity, the coframe $e^a$and Lorentz connection $\omega^{ab}$combine into a Cartan connection valued in the Poincaré algebra. Although torsion vanishes microscopically in vacuum GR, the holonomy of the Cartan connection around a loop $\gamma$enclosing the radiative region,

$$\mathcal{U}(\gamma) = \mathcal{P}\exp \oint_{\gamma} ( \omega^{ab} J_{ab} + e^a P_a),$$

need not return to the identity after a gravitational-wave burst. Its translational and rotational components correspond precisely to displacement memory and spin memory, providing a geometric interpretation of memory as a finite Poincaré holonomy between the initial and final vacuum states.

**B.2 Relation to BMS Symmetry**

At null infinity, the asymptotic symmetry group is the BMS group,

$$\mathrm{BMS} = \mathrm{Lorentz} \ltimes \mathrm{Supertranslations}.$$

A gravitational-wave burst interpolates between inequivalent BMS vacua. The change in Bondi shear,

$$\Delta C_{AB} = \int N_{AB}\, du,$$

encodes the displacement memory, and since$E_{AB} = -\frac{1}{2}\partial_u N_{AB} + O(r^{-1})$, we obtain

$$\Delta E_{AB} \propto \Delta C_{AB}.$$

In the bulk formalism developed here, the effective distortion satisfies

$$\beta_{ij}^{\mathrm{eff}} \propto \Delta E_{ij},$$

so that the Burgers vector

$$b_i = \oint \Delta h_{ij}^{TT} dx^j$$

becomes the bulk representative of the BMS supertranslation charge shift

$$b_i \ \leftrightarrow \ \Delta Q_{\mathrm{supertranslation}}.$$

**B.3 Burgers Vector and BMS Vacuum Degeneracy**

In defect theory, the Burgers vector measures the failure of closure of a transported frame. Asymptotically, a supertranslation $u \to u + f(\theta, \phi)$shifts the Bondi coordinates and selects a different BMS vacuum. Since displacement memory is equivalent to $\Delta C_{AB} \sim \Delta h_{ij}^{TT}$, the loop integral $\oint \Delta h_{ij} dx^j$encodes exactly this vacuum transition. Thus:

- BMS vacuum degeneracy ↔ translational non-closure,
- Supertranslation charge shift ↔ Burgers vector,
- Memory ↔ finite translational holonomy.

**B.4 Effective Torsion as a Bulk Density**

In gauge-theoretic language, torsion is the field strength associated with translations. In the effective picture used here,

$$T_{\mathrm{eff}}^a = d(\Delta e^a), \Delta e^a \sim \Delta h,$$

the effective torsion is therefore the bulk density associated with the integrated supertranslation transition. Although GR remains microscopically torsion-free, $T_{\mathrm{eff}}$provides a convenient coarse-grained encoding of the defect-like structure left by gravitational-wave memory — the spacetime analogue of a dislocation in continuum mechanics.